\documentclass[english, aps]{revtex4}%russian,
\usepackage{graphicx}
\usepackage{amsmath}
\begin{document}
%\date{\today}

\title{Regularization of the Coulomb scattering problem }

\author{V. G. Baryshevskii, I. D. Feranchuk and P. B. Kats}

\address{Byelorussian State University, F.Skariny Av., 4, 220050 Minsk,
Republic of Belarus}

\begin{abstract}
Exact solutions of the Schr\"odinger equation for the Coulomb potential are used in the scope of both stationary
and time-dependent scattering theories in order to find the parameters which define regularization of the
Rutherford cross-section when the scattering angle tends to zero but the distance r from the center remains
fixed. Angular distribution of the particles scattered in the Coulomb field is investigated on the rather large
but finite distance r from the center. It is shown that the standard asymptotic representation of the wave
functions is not available in the case when  small scattering angles are considered.  Unitary property of the
scattering matrix is analyzed and the "optical"  theorem for this case is discussed. The total and transport
cross-sections for scattering of the particle by the Coulomb center proved to be finite values and are
calculated in the analytical form. It is shown that the considered effects can be essential for the observed
characteristics of the transport processes in semiconductors which are defined by the electron and hole
scattering in the fields of the charged impurity centers.
\end{abstract}

\maketitle


\section{Introduction}

Scattering of non-relativistic charged particles by the Coulomb
center is one of the  canonical problems both in classical and
quantum mechanics which is known as the Rutherford problem. It is
the standard point of view  that the differential cross-section
$d\sigma (\theta)$ of the particle scattering to the solid angle
$d\Omega$ has the same form in the both cases (for example, Refs.
\cite{landau1}, \cite{newton})

\begin{eqnarray}
\label{1} d\sigma (\theta) = \sigma(\theta)  d\Omega =
(\frac{\alpha}{2 m v^2})^2 \frac{d\Omega}{\sin^4 \theta/2}.
\end{eqnarray}

\noindent Here  m and $v$ are the particle mass and velocity
correspondingly, parameter $\alpha$ defines the amplitude of the
Coulomb potential $U(r) = \alpha/r$.

So, the main measured characteristic of the scattering process in
the Coulomb field has the non-integrable singularity in the limit
$\theta \rightarrow 0$ (in quantum theory the singularity  exists
also in the scattering amplitude). Fortunately, this singularity
doesn't lead to any problem when describing of the most real
experiments because particles are scattered by the systems with
zero total charge. In this case the singularities conditioned by
the scattering centers of opposite signs are compensated and the
cross-section proves to be regular in the entire angular range.
Nevertheless, there are some physical systems where one should
consider the problem of regularization when calculating such
integral scattering characteristics as the total $\sigma_{tot}$
and transport $\sigma_{tr}$ cross-sections

\begin{eqnarray}\label{2} \sigma_{tot} = \int  d\sigma (\theta), \quad \sigma_{tr} = \int (1 - \cos \theta)
 d\sigma (\theta).
\end{eqnarray}

As for example, we can mention calculation of the characteristics
of kinetic processes in plasma and impurity semiconductors or
collisions of the charged particles in beams. In such cases one
should introduce some phenomenological parameter $\theta_{min}$
for cutting off the cross-section (\ref{1}) with angles $\theta <
\theta_{min}$. This parameter can be defined by various physical
reasons. Particularly, in the framework of the classical mechanics
the small angle scattering is defined by the particles with large
impact parameter \cite{landau2} connected with a long range
character of the Coulomb potential. Therefore, the small angle
cone can be excluded from the consideration because of finite
transversal width $a$ of the incident beam with $\theta_{min}\sim
a/r$ \cite{sing}.

Another approaches are used when the mobility of the charge
carriers is calculated in the impurity semiconductors. The models
of Brooks-Herring \cite{herring} and Conwell-Weisskopf
\cite{conwell} are mostly used for this problem at present. These
models correspond to different ways for estimation of the
parameter $\theta_{min}$ connected with screening of the Coulomb
potential. However, such estimations have only qualitative
character and some additional phenomenological parameter should be
introduced for more precise description of the mobility as it was
shown recently in the paper \cite{poklonskii}. Accurate
calculation of the integral values characterizing the charge
carrier scattering by impurities is actual because of high
accuracy of measurement of these values in real semiconductors
(for example, \cite{semiconducter}). Solution of this problem is
of great interest also for analysis of the electron transport in
nanostructures such as quantum wires \cite{wire}, superlattices
and films \cite{film}, nanotubes \cite{tube}.

Regularization problem for the Coulomb cross-section is
essentially more principal  in the framework of the quantum
theory. The matter is that the exact wave function  for the states
of  the continuous spectrum is well known  \cite{landau1} and it
has no any singularity even in the  case of the plane incident
wave which corresponds to the beam with the infinite  transversal
width.  It should mean that the singularity of the scattering
amplitude is not intrinsic feature of the Coulomb system in the
scope of quantum mechanical description. Possibly, it could be
conditioned by not completely adequate interpretation of the
asymptotic behavior of the wave function in this case. One can
expect that some characteristic, or "kinematical", regularization
parameter $\theta_{0}$ should exist which doesn't connect with the
initial state of the system unlike the value  $\theta_{min}$. In
general case the regularized cross-section should depend on both
parameters.

It is essentially to emphasize that some specific characteristics
of the Coulomb scattering problem have been widely discussed in
monographs and textbooks. As for example, it was shown in the book
\cite{sing} that the connection between the impact parameter and
scattering angle becomes indefinite in the case of $\theta = 0$,
therefore the scattering cross-section for zero angle can't be
calculated in classical dynamics. It is also well known that
long-range character of the Coulomb potential leads to the
logarithmic distortion of phase in the asymptotic form of the wave
function (for example  \cite{landau1}). However, the problem of
the cross-section regularization has not been considered in these
discussions.

This question was analyzed for the first time in our paper
\cite{1971}. It was shown that the standard asymptotic
representation of the wave function was not actually formed in the
range of small angles when considering the scattering processes by
the long-range potentials  ($U(r) \sim 1/r^s; s \leq 3$). In the
result the canonical definition of the scattering amplitude proved
to be unavailable. Born approximation over the potential $U(r)$
and the non-stationary collision theory \cite{goldberger} were
used in our work \cite{1971} in order to calculate the scattering
cross-section without any singularities. We can also mention
several papers ( \cite{zack} and references therein) where it was
shown that the interference between incident and scattered waves
changed the asymptotic form of the wave function and could be
essential in real experimental conditions even in the case of some
short-range potentials.

In the present paper we consider the non-asymptotic analysis of
the observed characteristics for the non-relativistic Coulomb
scattering problem  out of the framework of the perturbation
theory. We use the exact solutions of the Schr\"odinger equation
in order to answer the following questions: 1)which "intrinsic"
kinematical parameter defines regularization of the Rutherford
cross-section in the framework of the stationary scattering
theory; 2) how does this regularization depend on "external"
parameters such as the transversal width of the incidence wave
packet or  effective cutting off of the potential; 3) which way
can one calculate non-asymptotic values for the integral
scattering characteristics $\sigma_{tot}, \sigma_{tr}$; 4)what is
the analog of the "optical" theorem $4 \pi Im f(0) = k
\sigma_{tot}$ \cite{landau1}, \cite{newton} in the case of the
Coulomb potential? It seems to us that the answers for these
questions have the important methodical value for understanding
the scattering processes in the field of long-range potentials but
have not been discussed earlier. Besides, these results can be
also essential for some applications such as the above-mentioned
transport processes in the semiconductors with the charged
impurities.

The paper is organized as follows. In Sec. 2 the differential
scattering cross-section is defined without asymptotic
representation of the wave functions  and the kinematical
regularization parameter is found for the Rutherford problem. The
most important integral characteristics of the scattering problem
are calculated in Sec.3. In Sec.4 the scattering operator and the
conservation of the total flux are analyzed. The time-dependent
consideration of the collision process is discussed in Sec. 5 and
influence of the incident beam parameters and screening of the
potential to the observed scattering characteristics is estimated.
The scattering characteristics of the carriers  in non-degenerated
semiconductors with  the charged impurities are calculated in Sec.
6 and the results are compared with the experimental values of the
carrier mobility in real systems.

\section{Non-asymptotic calculation of the differential cross-section for the Coulomb scattering}

Let us remind the standard definitions of the scattering theory in
the stationary quantum mechanics. It is well known \cite{landau1},
that in this case the wave functions of the continuous spectrum
$\psi_{\vec{k}}(\vec{r})$ should be found as the solutions of the
Schr\"odinger equation

\begin{eqnarray}
\label{3} [ - \frac{\hbar^2}{2m}(\Delta + k^2) + U(\vec{r})
]\psi_{\vec{k}}(\vec{r}) = 0,
\end{eqnarray}

with the following asymptotic boundary conditions (Fig.1 shows all
necessary notations).
\begin{figure} [h]
\includegraphics[scale=0.8]{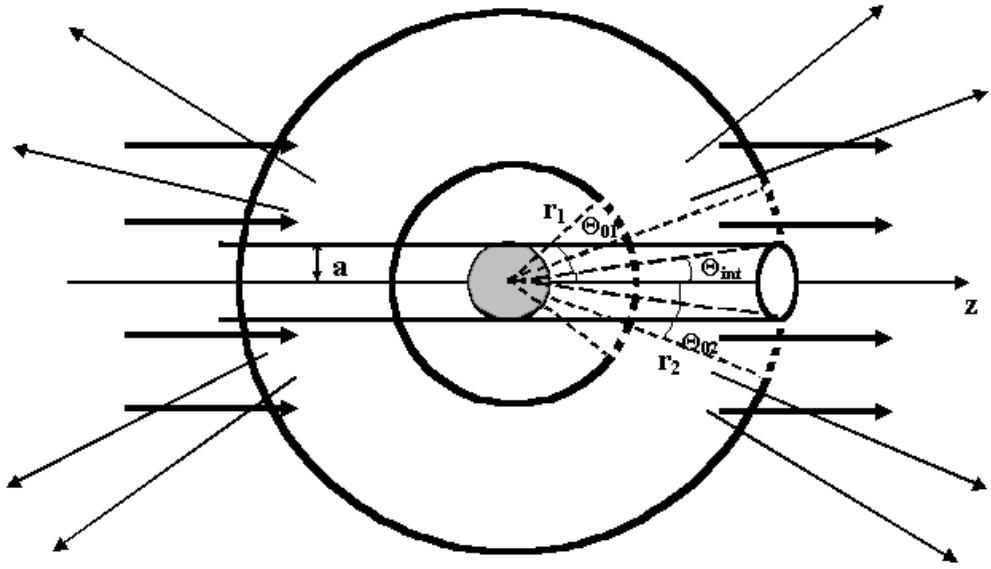}%width = 16cm,height =9cm
\caption{Sketch of the scattering process in the stationary case.}
\end{figure}

\begin{eqnarray}
\label{4} \psi_{\vec{k}}(\vec{r}) \sim e^{i \vec{k}\vec{r}}, \quad
\vec{k}\vec{r} \rightarrow  -\infty;
\end{eqnarray}

\begin{eqnarray}
\label{5} \psi_{\vec{k}}(\vec{r}) \sim e^{i \vec{k}\vec{r}} +
f(\theta) \frac{e^{ikr}}{r}, \quad r \gg R.
\end{eqnarray}

Here $\vec{k}$ is the wave vector; the value  R defines characteristic radius of the potential action  with the
center point $r=0$ ($R \rightarrow \infty$ in the case of the Coulomb field); the wave function is supposed to
be normalized to one particle, so that the flux density in the incident state  is:

\begin{eqnarray}
\label{6a} \vec{j} = \frac{\hbar}{2 i m} [\psi_{\vec{k}}(\vec{r})^* \nabla \psi_{\vec{k}}(\vec{r}) -
\psi_{\vec{k}}(\vec{r}) \nabla \psi_{\vec{k}}(\vec{r})^*] \simeq \frac{\hbar}{m}\vec{k} \equiv \vec{j}_0, \quad
\vec{k}\vec{r} \rightarrow  -\infty;
\end{eqnarray}

The flux density in the asymptotic state (\ref{5}) is divided on
two components:

\begin{eqnarray}
\label{7a} \vec{j} = j_l\frac{\vec{k}}{k} + j_{sc}\frac{\vec{r}}{r}, \quad r \gg R.
\end{eqnarray}

One of them $j_l$ (longitudinal component) corresponds to the
particles passed through the field without interaction and the
second one $j_{sc}$ (radial component) describes the scattered
particles. It leads to the standard definition of the
cross-section:

\begin{eqnarray}
\label{8a} d\sigma (\theta) = \frac{j_{sc}}{j_0} r^2 d\Omega; \nonumber\\ \sigma (\theta) = |f(\theta)|^2.
\end{eqnarray}

It should be noted that the longitudinal flux is also changed $j_l
< j_0$, and its decrease is defined by the total scattering
cross-section in accordance with the "optical" theorem
\cite{landau1}.

Evidently, that the definition (\ref{8a}) is based essentially on
the  asymptotic regime (\ref{5}) for the wave function in the
observation point r. Accordingly to the terminology used in
radio-physics and optics (for example, \cite{born}), it means that
the particle should go out of the "near" zone, where the action of
the potential is still essential, and pass to the "far", or
"wave", zone. The boundary between these zones is defined by the
condition that the interference between incident and scattered
waves becomes negligible that is the difference between their
phases satisfies the inequality

\begin{eqnarray}
\label{9}
k r - \vec{k} \vec{r} = 2 k r \sin^2 (\theta/2) > 1; \nonumber\\
\theta > \theta_0 \equiv \sqrt{\frac{2}{kr}}.
\end{eqnarray}

We suppose further that for all real collisions the condition $k r
\gg 1$  is fulfilled.

It is clear that the boundary of the "wave" zone depends both on the distance r from the center and the
scattering angle  $\theta$ (Fig.1). It means that in general case there is a part of the particle flux which can
not be described by the asymptotic wave function (\ref{5}) even for rather large distance r. Certainly, that
this property does not depend on the radius of the potential auction. However, the question is: what is the
contribution of these particles to the integral scattering process? When the distance from the center r is
fixed, the number of particles scattered to the "near" zone  $\theta < \theta_0$ can be estimated as

\begin{eqnarray}
\label{10} N_{dif} \simeq j_0 \sigma(0) \theta^2_0 \sim \frac{j_0
\sigma(0)}{k r},
\end{eqnarray}

The cross-section $\sigma(0)$ is restricted for the potentials
with the finite action radius R, therefore the $N_{dif}$ decreases
quickly at the large distance. It means that the contribution of
these particles to the observed scattering characteristics is
negligible for the most real experiments. The detailed analysis of
the "near" and "wave" zone formation for the scattering problem
with the short-range potential has been recently considered in the
paper  \cite{zack}.

The picture changes fundamentally in the case of the long-range potential $(R \rightarrow \infty)$ . The value
$N_{dif}$ can even increase with the distance and its contribution to the formation of the scattering flux can
be essential. Particularly, the analogous estimation in the case of the Coulomb field leads

\begin{eqnarray}
\label{10a} N_{dif} \gg j_0 (\frac{\alpha}{ m
v^2})^2\frac{4}{\theta^2_0} \sim k r.
\end{eqnarray}

It means that the asymptotic boundary condition (\ref{5}) is not
available in the entire range of the scattering angles and the
nonasymptotic expression for the wave function should be used in
the case of small angles. It is important to stress that this
circumstance does not connect with the width of incident beam and
defines by the characteristic feature of the potential itself.

So, the considered regularization problem for the Rutherford cross-section in the scope of the stationary
scattering theory is reduced to the analysis of the space flux distribution on the basis of the well known exact
solution of the equation (\ref{3}) with the potential $U(r) = Z e^2/r$ but without handling to the asymptotic
representation of the wave function.

We will use the following form of the  normalized wave function
\cite{landau1}

\begin{eqnarray}
\label{11} \psi_{\vec{k}}(\vec{r}) = N e^{i\vec{k}\vec{r}} F [\pm
i\xi, 1, i (k r - \vec{k}\vec{r})];
\quad N = e^{\pm \frac{\pi}{2}\xi}\Gamma (1 \mp i \xi);\nonumber\\
\xi = \frac{\alpha }{\hbar v}; \quad \alpha = Z e^2; \quad v =
\frac{\hbar k}{m},
\end{eqnarray}

where $F(a,b,t)$ is the confluent hypergeometric function; $\Gamma
(t)$ is the Gamma - function; the upper sign in the formulas
corresponds to the attraction field and the lower one corresponds
to the repulsion potential.

Let us show that the flux density in the formula (\ref{6a})
calculated with the exact wave function can also be divided by two
components in accordance with the formula (\ref{7a}) as it was in
the asymptotic regime  . For this purpose one can use the
following representation of the function F as the superposition of
two confluent hypergeometric functions of the 3-rd genus $U_{1,2}
(a,b,t)$ \cite{mors}

\begin{eqnarray}
\label{11a} F (\pm i\xi, 1, i z) = \frac{1}{\Gamma(\pm i\xi)}U_1
(\pm i\xi, 1, i z) + \frac{1}{\Gamma(1\mp i\xi)}U_2
(\pm i\xi, 1, i z);\nonumber\\
U_1 (\pm i\xi, 1, i z) = (z)^{\pm i \xi}\frac{e^{i z}}{\Gamma (1
\mp i \xi)}e^{\mp \pi \xi} G_1(\pm i\xi, i z), \nonumber\\
G_1(\pm i\xi, i z) = \int_{0}^{\infty} e^{-u} u^{\mp i \xi} (1 -
\frac{u}{i z})^{\pm i \xi} \frac{du}{i z - u} ;
\nonumber\\
U_2 (\pm i\xi, 1, i z) = (z)^{\mp i \xi}\frac{1}{ \Gamma (\pm i
\xi)}e^{\mp \pi \xi} G_2(\pm i\xi, i z),
\nonumber\\
G_2(\pm i\xi, i z) = \int_{0}^{\infty} e^{-u} u^{\pm i \xi - 1} (1 + \frac{u}{i z})^{\mp i \xi  } du; \nonumber\\
z = (k r - \vec{k}\vec{r}) = k r (1 - \cos \theta).
\end{eqnarray}

Let us also mention the connection between these functions and the
confluent hypergeometric functions of the 2-nd genus $U(a,b,t)$
\cite{nikiforov}

\begin{eqnarray}
\label{11b} U_1 (a, b, t) = U(b-a, b, -t)e^{t} e^{\pm i \pi(a-b)},
\nonumber\\ U_2 (a, b, t) = U (a, b, t)e^{\pm i a \pi},\nonumber\\
U (a, b, t) = \frac{1}{\Gamma(a)}\int_{0}^{\infty} e^{-t u} u^{a - 1} (1 + u)^{b-a-1} du; .
\end{eqnarray}

When the Rutherford cross-section is calculated by means of the
standard definition  this representation permits one to find the
asymptotic form of the wave function in the limit  $z \gg 1$
\cite{landau1}. This case corresponds to the "wave" zone when the
function $U_1$ transforms to the spherical wave  and the function
$U_2$ tends to the plane wave. However, both these functions are
well defined also in the "near" zone ($z < 1$) when they can be
calculated by means of the following serieses  \cite{mors}:

\begin{eqnarray}
\label{11c} U_1 (\pm i\xi, 1, i z) = \frac{1}{2}\Gamma(\pm i\xi) \{ F (\pm i\xi, 1, i z) + \nonumber\\
\frac{e^{\mp 2\pi \xi}- 1}{2\pi i}[ [2 \ln(i z) \mp i \pi \coth
(\pi \xi)- i\pi + 2 \psi (\pm i\xi)] F (\pm i\xi, 1, i z) +
\nonumber\\2 \sum_{m=1}^{\infty} \frac{\Gamma (m \pm i \xi)
}{\Gamma (\pm i \xi)(m!)^2}[\psi (m
\pm i \xi) - \psi (\pm i \xi) + 2\psi(1) - 2 \psi (m+1)] (i z)^m ] \};\nonumber\\
U_2 (\pm i\xi, 1, i z) = \frac{1}{2}\Gamma(1 \mp i\xi) \{ F (\pm i\xi, 1, i z) - \nonumber\\
\frac{e^{\mp 2\pi \xi}- 1}{2\pi i}[ [2 \ln(i z) \mp i \pi \coth
(\pi \xi)- i\pi + 2 \psi (\pm i\xi)] F (\pm i\xi, 1, i z) +
\nonumber\\2 \sum_{m=1}^{\infty} \frac{\Gamma (m \pm i \xi)
}{\Gamma (\pm i \xi)(m!)^2}[\psi (m \pm i \xi) - \psi (\pm i \xi)
+ 2\psi(1) - 2 \psi (m+1)] (i z)^m ] \},
\end{eqnarray}

where $\psi (t)$ is the logarithmic derivative of the
Gamma-function.

When the representation  (\ref{11a}) is used in the formula (\ref{6a}) one should take into account only the
derivatives from the exponents  because the conditions $k r \gg 1; \ z \sim 1$ supposed to be fulfilled. As for
example,

$$
- i (\frac{\vec r}{r}\vec{\nabla})[ e^{ikr} G_1 (i z)] = e^{ikr}[k
G_1 + \frac{z}{r}G'_1(iz)] \simeq e^{ikr}k
 G_1[1 + O (\frac{1}{k r})].
$$

This representation permits one to find the scattering flux $j_{sc}$ directed to the observation point along the
vector $\vec{r}$  without use of  the asymptotic form (\ref{5}) of the wave function. In the result the
scattering cross-section can be defined in the entire range of the angles $\theta$ in the following form:

\begin{eqnarray}
\label{12} \sigma_1(\theta)d\theta  =\sin \theta r^2\int_{0}^{2\pi} d \varphi \frac{j_{sc}}{j_0}d\theta =
\nonumber\\
2\xi \sinh (\pi \xi)e^{\mp \pi \xi}|G_1 (\pm i \xi, i z)|^2 r^2 \sin \theta d\theta.
\end{eqnarray}

One can see that the differential cross-section $\sigma_1(\theta)$
is finite for any angle in spite the function $G_1 (\pm i \xi, i
z)$ has the logarithmic singularity at zero angle as it follows
from the equation (\ref{11c}). But this value depends on the
distance between the center and observation point by non-trivial
way because of long-range action of the potential field to the
particle. The result of this action at small angles ("near"  zone)
does not reduce to varying the phase of the scattering amplitude
as it is takes place for the asymptotic range of angles ("wave"
zone)\cite{landau1}.

If one considers behavior of the function $G_1 (\pm i \xi, i z)$ in dependence on the scattering angle the
"kinematical" parameter $\theta_0$ for regularization of the Rutherford cross-section can be introduced by the
natural way. Actually, the asymptotic range of angles corresponding to the "wave" zone is defined by the
condition

\begin{eqnarray}
\label{13} z = k r - \vec k \vec r \simeq \frac{1}{2}k r \theta^2 \gg 1;\nonumber\\
\theta \gg \theta_0 = \sqrt{\frac{2}{k r}} \ll 1;\quad x \gg 1;\nonumber\\
x = \frac{\theta}{\theta_0}; \quad z \simeq x^2.
\end{eqnarray}

Here the dimensionless value x is introduced as the convenient variable for the angles compared with the width
of the "near" zone. Certainly, in the range of $x \gg 1$ the standard asymptotic representation of the integral
in the definition of the function $G_1 (\pm i \xi, i z)$ leads to the result corresponding to the formula
(\ref{1}) with new variable

\begin{eqnarray}
\label{14} \sigma_1(\theta) \simeq 2 \pi(\frac{\xi}{k })^2
\frac{\sqrt{2} (k r)^{3/2}}{x^3} = 8 \pi (\frac{ \alpha}{m v^2})^2
\frac{1}{\theta^3}; \quad x \gg 1.
\end{eqnarray}

It is well known that the interference between the scattering flux
and the flux directed along the initial velocity of the particle
does not take into account in scope of any quantum scattering
theory based on the solutions of the stationary Schr\"odinger
equation \cite{landau1}. So, in order to use the formula
(\ref{12}) in the range $x < 1$ corresponding to the "near" zone,
one should compare it with the angle width of the zone where the
above mentioned interference is still essential. It is clear that
the angle width of such "interference" zone does not depend on the
dynamics of the interaction between the particle and field. It is
defined only by the transversal width $a$ of the incident particle
wave packet (Fig.1). One can estimate the angle width
$\theta_{int}$ of the "interference" zone as follows

\begin{eqnarray}
\label{15} \theta_{int} \simeq \frac{a}{r}.
\end{eqnarray}

It means that one can consider the scattering flux in the near zone and at the same time neglect by its
interference with the incident beam if the following conditions are fulfilled

\begin{eqnarray}
\label{15a} \theta_{int}  < \theta < \theta_0 = \sqrt{ \frac{2}{k
r}}; \quad \frac{k a^2}{r} \ll 1.
\end{eqnarray}

These inequalities are satisfied in the case of rather large $r$
as it usually supposed in the scattering theory. More accurate
analysis of this factor will be considered below (Sec.3) in the
framework of the time-dependent theory of collisions
\cite{goldberger}. But one can estimate just now the contribution
of the "interference" zone to the integral scattering
characteristics which are finite values in our consideration
unlike the asymptotic analysis. As for example, the ratio of the
particle flux scattered at the "interference" and  "near" zones
can be estimated as

\begin{eqnarray}
\label{16} \delta = \frac{j_{int}}{j_{dif}} \simeq
\int_{0}^{\theta_{int}}\sigma_1(\theta) d \theta/
\int_{0}^{\theta_{0}}\sigma_1(\theta) d \theta \simeq  \frac{k
a^2}{2r} \ll 1.
\end{eqnarray}

It remains small under standard conditions of the collision theory
\cite{goldberger} and one can analyze distribution of the flux
density in the "near" zone neglecting its interference with the
incident flux. It permits one to find the leading terms of the
differential scattering cross-section at small angles using the
series  (\ref{11c})

\begin{eqnarray}
\label{17} \sigma_1(\theta) \simeq 8 \sqrt{2} \xi e^{\mp \pi \xi}\sinh (\pi \xi) x (\ln x)^2
\frac{r^{3/2}}{\sqrt{k}}.
\end{eqnarray}

Fig.2 compares the accurate and asymptotic scattered fluxes for
various values of the variable $x$ and parameter $\xi$. It is
interesting to pay one's attention to the essentially different
behavior of the nonasymptotic flux in "near" zone for scattering
by the attractive and repulsive centers in contrast to the
Rutherford cross-section (\ref{1}) which is independent of the
potential sign for any value $\xi$. One can see that the
regularized differential cross-section  (\ref{12}) in the "near"
zone is essentially non-invariant relatively to the sign of the
charge if the parameter  $\xi \geq 1$. It should be noted that the
effect of slightly different interaction of the charge carriers
with the impurities of different signs is well known in the
semiconductor physics. It is usually considered there by means of
the Friedel sum rule \cite{fridel}  using the partial expansion of
scattering amplitude in the series of orbital momenta.

As it follows from Eq. (\ref{17}), the scattering flux in the case
of attractive potential varies rather slowly with increase of the
parameter $\xi$, but it grows  exponentially in the case of
repulsion. Certainly, such behavior of the cross-section takes
place only in the narrow angle domain (\ref{15a}) and compensates
the exponential decrease of the flux just along the line $\theta =
0$ which is well known for the repulsive potential \cite{landau1}

\begin{figure} [h]
%\begin{minipage}{0.1\texwidth}\centering
\includegraphics[width = 5cm,height =5cm]{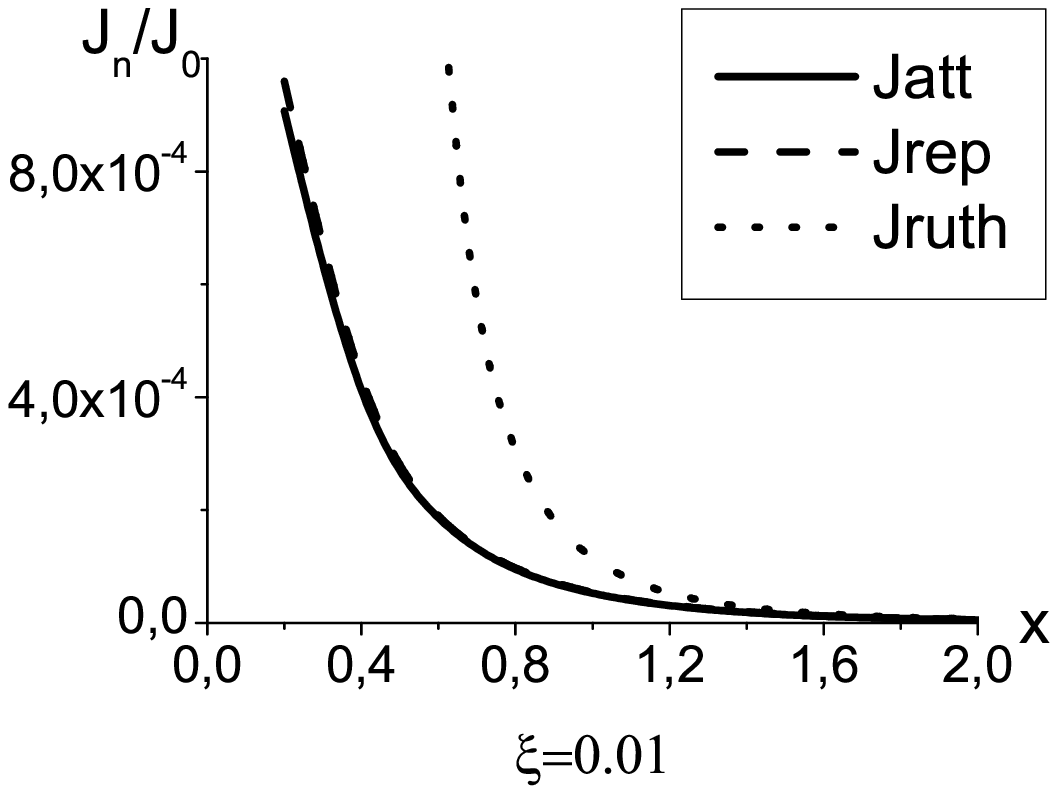}
%\end{minipage}\hfill\
%\begin{minipage}{0.1\texwidth}\centering
\includegraphics[width = 5cm,height =5cm]{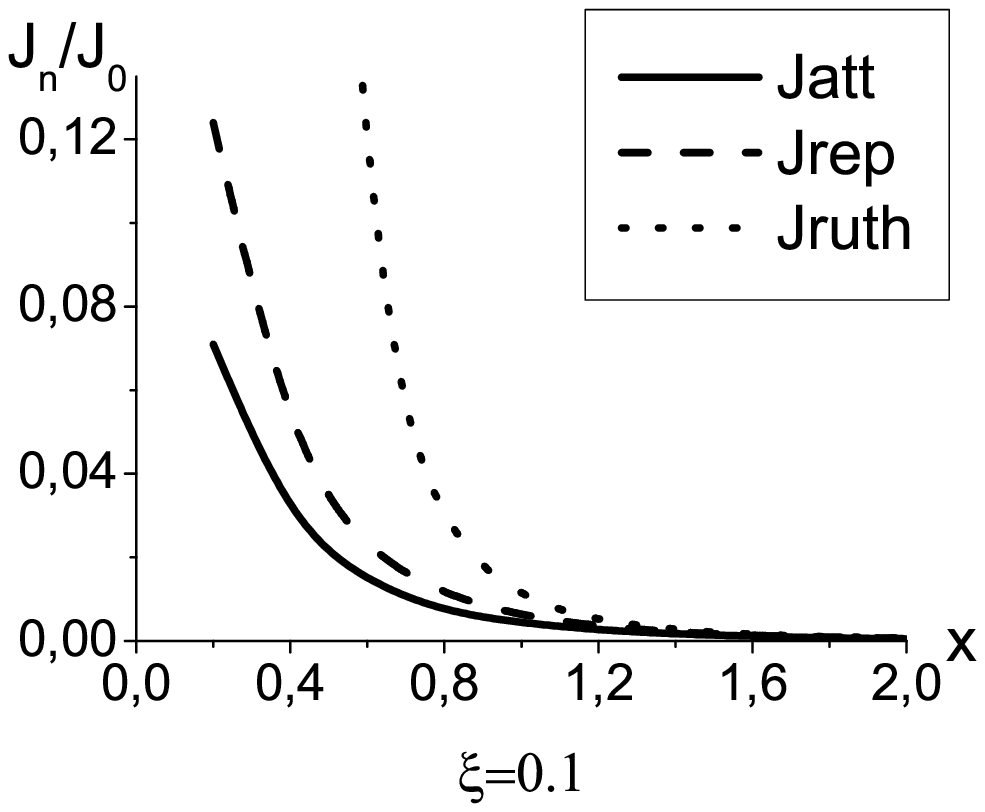}
%\end{minipage}\hfill\
%\begin{minipage}{0.1\texwidth}
%\centering

\includegraphics[width = 5cm,height =5cm]{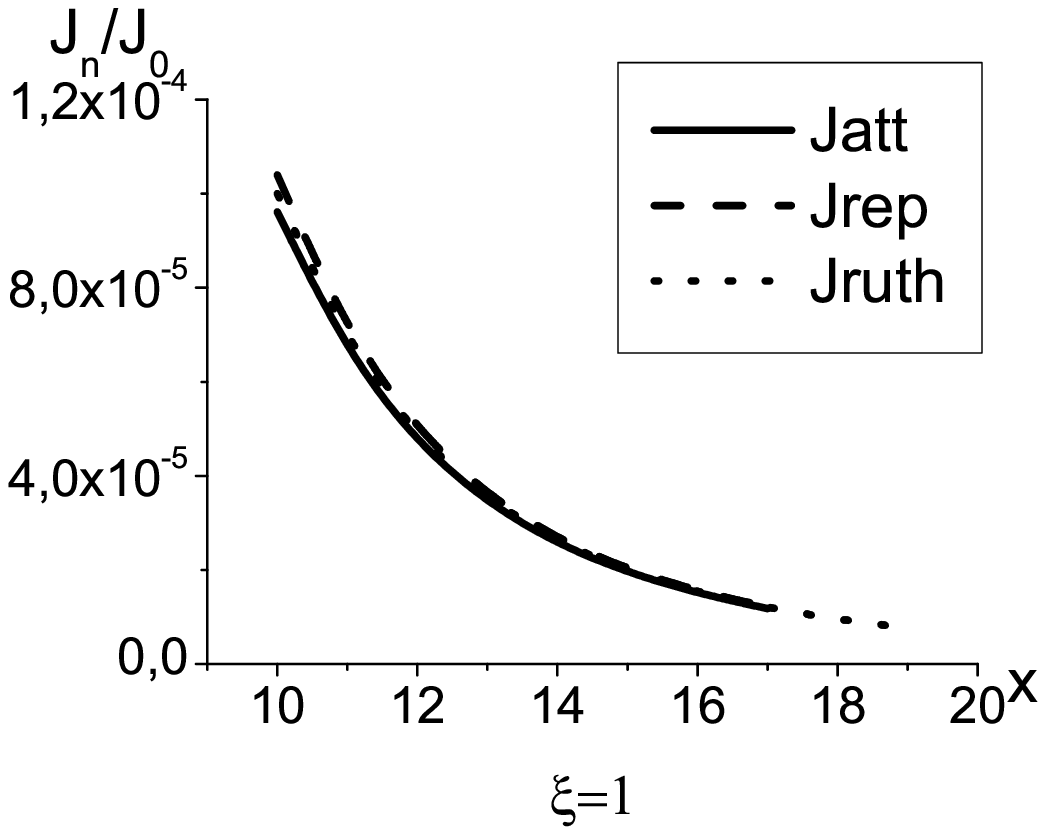}
%\end{minipage}\hfill\
\caption{ Ratio of the scattered flux $J_n$ to the density of the incident flux $J_0$. Solid line - the case of
attraction $J_{att}$, dashed line - the case of repulsion $J_{rep}$, dotted line - the Rutherford flux
$J_{ruth}$.}
\end{figure}

\section{Integral characteristics for the Coulomb scattering problem}

Let us now calculate the integral scattering characteristics for
the considered problem. In accordance with Eq. (\ref{12}) the
nonasymptotic expression for the total cross-section is defined by
the following integral

\begin{eqnarray}
\label{19} \sigma_{tot} = \int_{0}^{\pi} 2\xi \sinh (\pi \xi) |G_1 (\pm i \xi, i z)|^2 r^2 \sin \theta d \theta
.
\end{eqnarray}

One can use in this integral new variable $z$

\begin{eqnarray}
\label{20} \sigma_{tot} = \int_{0}^{2 k r} \xi \sinh (\pi \xi) |G_1 (\pm i \xi, i z)|^2 \frac{2r}{k}d z,
\end{eqnarray}

and represent it as the sum of two integrals

\begin{eqnarray}
\label{21a} \sigma_{tot} =\frac{2r}{ k}\xi \sinh (\pi \xi)\{ \int_{0}^{\infty}  |G_1 (\pm i \xi, i z)|^2 d z -
\nonumber\\\int_{2 k r}^{\infty}  |G_1 (\pm i \xi, i z)|^2 d z\} .
\end{eqnarray}

The asymptotic representation (\ref{11a}) for the function $G_1$
can be used in the whole interval of integration in the second
integral
$$
G_1 (\pm i \xi, i z) \simeq \frac{\Gamma (1 \mp i \xi)}{i z},
$$

and it leads to the following simple result

\begin{eqnarray}
\label{22} I_2 = \frac{2r}{ k}\xi \sinh (\pi \xi)\int_{2 k r}^{\infty}  |G_1 (\pm i \xi, i z)|^2 d z \simeq
\frac{\pi \xi^2}{k^2}.
\end{eqnarray}

Therefore it has the order of  $(k r)^{-1}$ in comparison with the
first integral and  its contribution to the total cross-section
can be omitted.  It permits one to find how the value
$\sigma_{tot}$ depends on the most essential parameters of the
problem

\begin{eqnarray}
\label{23} \sigma_{tot} =\frac{2\pi r}{k}\xi^2 I_{\pm} (\xi); \nonumber\\
I_{\pm}(\xi) = e^{\mp \pi \xi} \int_{0}^{\infty} |U_1 (1 \pm i
\xi,1, i z)|^2 d z .
\end{eqnarray}

Here we use again the canonical form for the confluent
hypergeometric function of the 2-nd genus\cite{mors}. The
universal functions  $I_{\pm}(\xi)$ depend only on the variable
$\xi$. They are defined by the converged integrals and can be
easily calculated numerically. Fig.3 shows the results of these
calculations.

\begin{figure} [h]
%\begin{minipage}{0.2\texwidth}\centering
\includegraphics[width = 5cm,height =5cm]{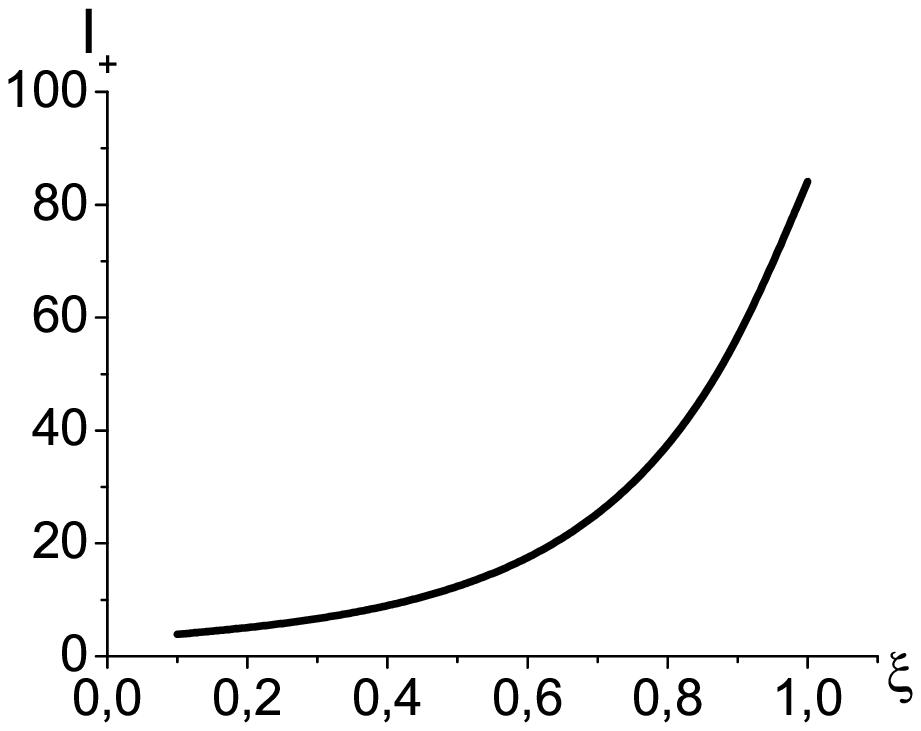}%
%\end{minipage}\hfill\
%\begin{minipage}{0.2\texwidth}\centering
\includegraphics[width = 5cm,height =5cm]{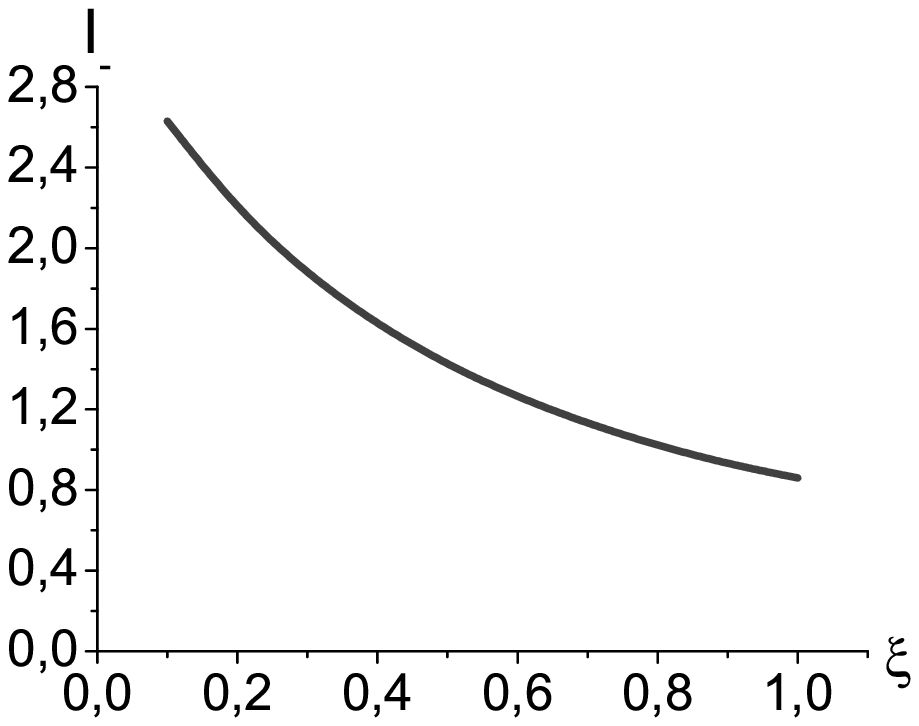}
%\end{minipage}\hfill\
\caption{Universal functions $I_{\pm}(\xi)$ which define the total
scattering cross-section as the functions of the parameter $\xi$
for the cases of repulsion and attraction.  }
\end{figure}

 Now let us consider
another integral characteristic of the scattering process, namely, the transport cross-section which is very
important value for a lot of applications. It is defined by the formula

\begin{eqnarray}
\label{24} \sigma_{tr} = \int_{0}^{\pi} 2 \xi \sinh (\pi \xi) |G_1
(\pm i \xi, i z)|^2 r^2 \sin \theta (1 - \cos \theta) d \theta .
\end{eqnarray}

If one uses the variable z in this integral and comes back to the
hypergeometric function of the 2-nd genus, Eq.(\ref{24})
transforms as follows

\begin{eqnarray}
\label{25} \sigma_{tr} =\frac{2\pi \xi^2}{k^2} e^{\mp \pi \xi} I^{tr}_{\pm} (\xi); \nonumber\\
I^{tr}_{\pm}(\xi) = \int_{0}^{2 k r} |U (1 \pm i \xi,1, i z)|^2 z
d z .
\end{eqnarray}

The integrand function for the transport cross-section is essentially suppressed in the range of small angles in
comparison with the total cross-section. Therefore  $\sigma_{tr}$ is defined by only the logarithm of the
distance to the observation point unlike to    $\sigma_{tot}$ proportional to this distance. Besides, this
function decreases rather slowly for the large $z$ and we can't use the trick analogous to Eq. (\ref{21a}) for
$\sigma_{tr}$. Nevertheless a series of transformations of the integral $I^{tr}_{\pm}(\xi)$  permits one to find
the analytical dependence on the coordinate $r$ with an accuracy of the order $(k r )^{-1}$. Let us separate the
integral on two parts by the following way
$$
I^{tr}_{\pm}(\xi) = \int_{0}^{1} |U (1 \pm i \xi,1, i z)|^2 z d z
+ \int_{1}^{2 k r} |U (1 \mp i \xi,1, i z)|^2 z d z.
$$

If one uses the asymptotic formulas for the hypergeometric
functions \cite{mors}

$$
|U (1 \pm i \xi,1, i z)|^2 \simeq \frac{e^{\pm \pi \xi}}{z^2} +
O[( k r )^{-3}],
$$

the second term in this integrals is transformed identically

$$
\int_{1}^{2k r}[ |U (1 \pm i \xi,1, i z)|^2 - \frac{e^{\pm \pi
\xi}}{z^2}]z d z + \int_{1}^{2 k r} \frac{e^{\pm \pi \xi}}{z} d z.
$$

The second integral here is calculated analytically but now the integrand expression in the first one decreases
rather quickly and the estimation analogous to  (\ref{21a}) can be used

\begin{eqnarray}
\label{26} \int_{1}^{\infty}[ |U (1 \pm i \xi,1, i z)|^2 -
\frac{e^{\pm \pi\xi}}{z^2}]z d z - \int_{2kr}^{\infty}[
|U (1 \pm i \xi,1, i z)|^2 - \frac{e^{\pm \pi \xi}}{z^2}]z d z \simeq \nonumber\\
\int_{1}^{\infty}[ |U (1 \pm i \xi,1, i z)|^2 - \frac{e^{\pm
\pi\xi}}{z^2}]z d z + O[( k r )^{-1}].
\end{eqnarray}

In the result the transport cross-section is defined by well
converged integrals

\begin{eqnarray}
\label{27} \sigma_{tr} = \frac{2\pi \xi^2 e^{\mp \pi\xi}}{k^2}
\{\int_{0}^{1}|U (1 \pm i \xi,1, i z)|^2 z d z
+ \nonumber\\
\int_{1}^{\infty}[ |U (1 \pm i \xi,1, i z)|^2 -  \frac{e^{\pm \pi
\xi}}{z^2}]z d z + e^{\pm \pi \xi} \ln (2 k r)\} .
\end{eqnarray}

Fig.4 shows the results of numerical calculation of the universal functions $I^{tr}_{\pm}(\xi) = [k^2
\sigma_{tr} - 2\pi \xi^2\ln (2 k r)]$.

\begin{figure} [h]
\includegraphics[width = 7cm,height =7cm]{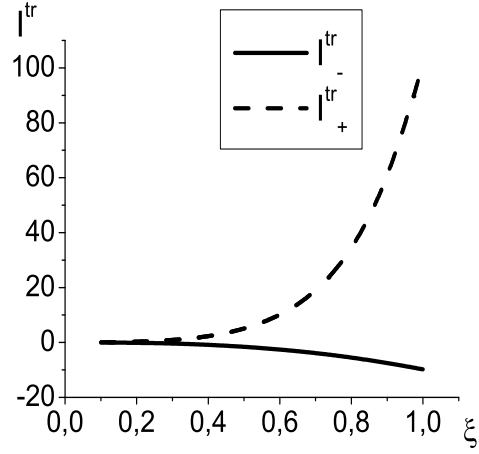}
\caption{ Universal functions $I^{tr}_{\pm}(\xi)$ for the
transport cross-section in the cases of attraction and repulsion.
Solid line - the case of attraction, dashed line - the case of
repulsion }
\end{figure}

\section{Scattering operator and conservation of the flux in scope of the stationary theory}

As it follows from the results of the preceding section the
integral scattering characteristics calculated on the basis of
nonasymptotic consideration increase together with distance $r$ to
the observation point. It seems for the first sight that it can
contradict to the conservation of the total flux of the particles
when $r$ becomes rather large. However, let us show that this
dependence expresses only the fact that the potential influences
on the scattering process at any distance from the center but the
scattering flux remains essentially less than integral incidence
flux at any $r$. For qualitative analysis one should take into
account that in scope of the stationary scattering theory the
quantum state of the incident particle is described by the plane
wave. Then the total incidence flux $J_0$ through the sphere with
radius $r$ corresponding to the observation point can be estimated
as follows

$$
J_0 \simeq j_0 \pi r^2.
$$

Then the ratio of the scattering and incident integral fluxes is

\begin{eqnarray}
\label{28} \frac{j^{tot}_{sc}}{j_0} \simeq
\frac{\sigma_{tot}}{r^2} = \frac{2\pi}{k r}\xi^2 I_{\pm} \ll 1.
\end{eqnarray}

It is also important to consider this problem more precisely. It
is known \cite{landau1} that the condition of the total flux
conservation leads to the "optical" theorem in the quantum theory
of scattering by short-range potential when the amplitude of
scattering to the zero angle is the finite value. We use the same
approach \cite{landau1} in order to find the consequence of this
condition in the case of nonasymptotic analysis of the Coulomb
scattering.

Let us represent general solution of the Schr\"odinger equation in
the case of the elastic scattering as the linear combination of
the functions  (\ref{11}) with arbitrary coefficients $\Phi (\vec
n)$ which define the amplitudes of probability to find the state
with the wave vector $\vec k = k \vec n$ in the initial packet:

\begin{eqnarray}
\label{29} \Psi (\vec r)= \int \Phi (\vec n)
\psi_{\vec{k}}(\vec{r}) d \Omega_{\vec n} = N \int \Phi (\vec n)
e^{i k r\vec{n}\vec{n}'} F [\pm i\xi, 1, i k r (1 -
\vec{n}\vec{n}')] d\Omega_{\vec n} =
\nonumber\\
\int \Phi (\vec n) [(z)^{\pm i \xi}\frac{e^{i kr}}{\Gamma (\pm i
\xi)} G_1(\pm i\xi, i z) + (z)^{\mp i
\xi}\frac{e^{i k r\vec{n}\vec{n}'}}{ \Gamma (\pm i \xi)} G_2(\pm i\xi, i z)]d\Omega_{ \vec n}; \nonumber\\
\vec{n}' = \frac{\vec r}{r}; \quad z = k r (1 - \vec{n}\vec{n}'),
\end{eqnarray}
\noindent $d\Omega_{\vec n}$  is the element of the solid angle in
the direction of the vector  $\vec n$.

In accordance with the physical interpretation of the
contributions defined by the functions  $ G_1, G_2 $ to the total
wave function (\ref{11}), the first term in Eq.(\ref{29})
describes that part of the integral scattering operator
\cite{landau1} which corresponds to the formation of the
scattering wave. The term, proportional to the function  $ G_2 $,
describes the deformation of the wave packet conditioned by change
of the plane wave in the Coulomb field. One can estimate the
second term by the same method that was used for proof of the
"optical" theorem in the case of short-range potential
\cite{landau1} . If the condition $k r \gg 1$ is fulfilled, the
main contributions to this integral  are defined by the small
intervals near the points of the stationary phases when
integrating over  $\vec{n}$ . These points correspond to the
vectors  $\vec{n}_1 = - \vec{n}'$ and $ \vec{n}_2 = \vec{n}'$.
Near the first point the variable  $z \simeq 2 k r$ is very large.
Therefore one can use the asymptotic expression for the function
$G_2(\pm i\xi, i z)$ and the integrand has no singularities in
this case. In the result the contribution to the integral from the
domain close to this point defines the converged spherical wave
with the standard logarithmic distortion of its phase
\cite{landau1}

$$
\sim 2 \pi i \frac{ e^{- i k r \mp i \xi \ln 2 k r}}{ k r}\Phi (-
\vec{n}').
$$

When estimating the contribution to the integral from  the second
point of stationary phase corresponding to scattering at small
angles one should take into account that the function $G_2(\pm
i\xi, i z)$ has the logarithmic singularity in the point  $z = 0$.
Nevertheless, the rather smooth weight function $\Phi (\vec n)$
can be removed from the integral in the point  $\vec{n} =
\vec{n}'$. It leads to the following estimation:

$$
\sim 2 \pi \Phi ( \vec{n})e^{\mp \frac{\pi}{2}\xi}\frac{ e^{i k r
}}{ k r\Gamma (\pm i \xi)}[\int_{0}^{\infty} (z)^{\mp i \xi} e^{-i
z } G_2(\pm i\xi, i z)d z - \int_{2 k r}^{\infty} (z)^{\mp i \xi}
e^{-i z } G_2(\pm i\xi, i z)d z].
$$

The second integral in this expression can be omitted in the limit
$k r \gg 1$ and the initial wave function is represented in the
form:

\begin{eqnarray}
\label{30} \Psi (\vec r) \simeq 2 \pi i \frac{ e^{- i k r \mp i
\xi \ln 2 k r}}{ k r}\Phi (- \vec{n}') - 2 \pi i
\frac{ e^{i k r }}{ k r}[ A \Phi ( \vec{n}') + \int \hat f (\vec{n},\vec{n}') \Phi ( \vec{n})d \vec n]; \nonumber\\
A = \frac{i e^{\mp \frac{\pi}{2}\xi}}{\Gamma (\pm i \xi)}\int_{0}^{\infty} (z)^{\mp i \xi} e^{-i z } G_2(\pm i\xi, i z)d z;\nonumber\\
\hat f (\vec{n},\vec{n}') =\frac{i k r}{2 \pi\Gamma (\pm i
\xi)}e^{\mp \frac{\pi}{2}\xi} e^{ \pm i  \xi \ln k r}(1 -
\vec{n}\vec{n}')^{\pm i \xi} G_1[\pm i\xi, i k r (1 -
\vec{n}\vec{n}')].
\end{eqnarray}

It is more convenient to rewrite this expression in terms of the hypergeometric function of the 2-nd genus

\begin{eqnarray}
\label{31}
A = i e^{\mp \frac{\pi}{2}\xi}\int_{0}^{\infty} e^{-i z } U(\pm i\xi,1, i z)d z;\nonumber\\
\hat f (\vec{n},\vec{n}') =-\frac{i k r \Gamma (1\mp i \xi)}
 {2\pi\Gamma (\pm i \xi)}e^{\mp \frac{\pi}{2}\xi}  U[1\mp i\xi,1, -i
k r (1 - \vec{n}\vec{n}')].
\end{eqnarray}
 Now the function is represented as the superposition of the
ingoing and outgoing spherical waves and it permits one to
introduce the scattering matrix   \cite{Landau1} as the following
integral operator:

\begin{eqnarray}
\label{32} \hat S(\vec{n},\vec{n}') \simeq A
\delta_{\vec{n},\vec{n}'} + \hat f (\vec{n},\vec{n}').
\end{eqnarray}

Here $\delta_{\vec{n},\vec{n}'}$ is the unit operator which
corresponds to the wave passed without scattering and the
parameter A defines the change of its amplitude (in the case of
the short-range potential $A=1$ \cite{landau1}). The integral over
angles from the operator  $\int \hat f^*(\vec{n},\vec{n}')\hat f
(\vec{n}',\vec{n}) d \vec{n}'$ coincides exactly with the
expression for the total cross section (\ref{23}). Long-range
character of the potential is appeared in the fact that the
scattering matrix elements depend on the coordinate $r$. However,
it is very important to introduce such operator because just it
defines the kernel of the collision integral in the kinetic
equations for description of various transport processes
\cite{kinetic}. But if one uses such operator in the collision
integral for one-particle distribution function the additional
averaging over the coordinate should be fulfilled. Dependence of
the function $ f (\vec{n}',\vec{n}) $ on the coordinate is rather
smooth , therefore the value $r$ in this function can be
substituted as an average distance between the scattering centers
if the correlation between these centers can be neglected  (see
below $\S 6$). Analogous substitution was used in some well-known
models for regularization of the transport cross section of
scattering by the charged impurities in semiconductors
\cite{herring}, \cite{conwell}.

Unitary property of the matrix  $\hat S(\vec{n},\vec{n}')$  leads
to the "optical" theorem in the case of short-range potentials
\cite{landau1}. But if one uses this condition in the case of
Coulomb potential there  is the problem that the operator  $\hat f
(\vec{n},\vec{n}')$ has the logarithmic singularity in the limit
of coinciding arguments and one should define the way for
calculating  integral from the product of singular functions $\hat
f(\vec{n},\vec{n}')$ and $\delta_{\vec{n},\vec{n}'}$ in the
operator $\hat S \hat S^+ $ . Actually it means that the
asymptotic estimation of the integral in Eq.(\ref{30}) is
unavailable for the operator which is quadratic over the
scattering matrix. Therefore let us analyze separately the
conservation of flux considering the following integral

\begin{eqnarray}
\label{33} I =  \int d \vec S (\vec{\nabla} j (\vec r)) \equiv \int ((\vec{r}j (\vec r)) r d \Omega_{\vec{n}}; \nonumber\\
\vec j (\vec r) =  \frac{\hbar}{2 m i}\{\Psi^* (\vec
r)\vec{\nabla}\Psi (\vec r) - \Psi (\vec r)\vec{\nabla} \Psi^*
(\vec r)\},
\end{eqnarray}

\noindent with the total wave function (\ref{29}).

When the superposition (\ref{29}) is used in formula (\ref{33})
one can take into account the completeness of the coefficients
$\Phi (\vec n)$. Then integration over all directions in this
integral is equivalent to the integral from the flux
$\vec{j}_{st}$ calculated by means of the general formula
(\ref{33}) but with the stationary wave functions
$\psi_{\vec{k}}(\vec{r})$ defined by Eq. (\ref{11}) ( let us
consider the attractive potential for definiteness )

\begin{eqnarray}
\label{34} \vec j_{st} (\vec r) =  \frac{\hbar \pi \xi
e^{\pi\xi}}{m \sinh \pi \xi} \{\vec k |F [ i\xi, 1, i (k
r - \vec{k}\vec{r})]|^2 - \xi (\vec k - k \frac{\vec r}{r}) \Im (F F_1^*)\}; \nonumber\\
\vec{\nabla}F [ i\xi, 1, i (k r - \vec{k}\vec{r})] = \xi (\vec k -
k \frac{\vec r}{r}) F [ i\xi + 1, 2, i (k r - \vec{k}\vec{r})]
\equiv  \xi (\vec k - k \frac{\vec r}{r}) F_1.
\end{eqnarray}

Certainly, the value  $I$ is equal to zero identically because of
the flux conservation for the stationary scattering problem. The
"optical" theorem is followed from this condition if the
asymptotic form (\ref{5}) for the wave function can be used
\cite{landau1}. But in the considered problem this condition means
that the flux directed along the vector  $\vec k$ (it defines
change of the intensity of the incident wave), and the scattering
flux along the vector  $\vec r$ are connected as follows

\begin{eqnarray}
\label{35} \int d \Omega_{\vec{n}} (\vec k \vec n) |F [ i\xi, 1, i
(k r - \vec{k}\vec{r})]|^2 =  \xi \int d \Omega_{\vec{n}} (\vec k
\vec r - k ) \Im (F F_1^*).
\end{eqnarray}

As it was shown above the integrals over the angle for the Coulomb
scattering problem include essential contribution defined by
"near" zone. Therefore the both parts of Eq.(\ref{35}) depend on
the coordinate $r$ and the standard asymptotic expressions for
"optical" theorem is inapplicable because the total cross section
and the scattering amplitude at zero angle are tending to infinity
in this case. But if one shows that the leading terms of
Eq.(\ref{35}) are equal in the limit of large r ($ k r \gg 1$)  it
can be considered as the analog of the "optical" theorem for the
Coulomb potential.

In order to prove it let us use new variable for the integrals in
Eq. (\ref{35})

$$
z = k r - \vec{k}\vec{r}; \quad \sin \theta d \theta = \frac{d
z}{kr},
$$

\noindent and transform them as follows

\begin{eqnarray}
\label{36} \int_{0}^{2 k r}|F ( i\xi, 1, i z)|^2 d z = \int_{0}^{2
k r}\frac{z}{k r} \{|F |^2 + \xi \Im [ F(-i \xi, 1, - i z) F(i \xi
+ 1, 2, i z) ] \} d z.
\end{eqnarray}

One can estimate the integrals from the confluent hypergeometric
functions in the range $ k r \gg 1 $ by means of the following
approach. Integral in the left side of Eq.(\ref{36}) can be
transformed identically

\begin{eqnarray}
\label{37} J_1 = \int_{0}^{2 k r}|F ( i\xi, 1, i z)|^2 d z =
\lim_{\delta \rightarrow 0} [ \int_{0}^{\infty}|F |^2 e^{- \delta
z} dz  - \int_{2 k r}^{\infty}|F |^2 e^{- \delta z} d z ].
\end{eqnarray}

\noindent The parameter $\delta \rightarrow 0 $ is introduced for
the regularization of both integrals at upper limit. The
asymptotic form of the function F can be used in the second term
and the first term can be expressed through the hypergeometric
function $F(\alpha, \beta, \gamma, z)$ by means of the formula
(see, for example \cite{landau1})

\begin{eqnarray}
\label{38} J(\lambda) = \int_{0}^{\infty} e^{- \lambda z}
z^{\gamma - 1} F(\alpha, \gamma, k z) F(\alpha ',
\gamma, k' z) dz = \nonumber\\
\Gamma (\gamma) \lambda^{\alpha + \alpha ' - \gamma} (\lambda - k
)^{-\alpha}(\lambda - k' )^{-\alpha '} F [\alpha, \alpha ',
\gamma, \frac{k k'}{(\lambda - k )(\lambda - k' )}].
\end{eqnarray}

When the integrals from the functions with different second
arguments are calculated, the following recursion relation can be
used  \cite{mors}

$$
F(\alpha + 1, \gamma + 1, z) = \frac{\gamma}{z}[F(\alpha + 1,
\gamma , z) - F(\alpha , \gamma , z)].
$$

Let us write also the leading terms of the asymptotic expansions
for the functions  F è $F_1$ which are used for the integrals in
the limits  $(2 k r, \infty)$

$$
F ( i\xi, 1, i z) \simeq e^{-\pi\xi/2}[
\frac{z^{-i\xi}}{\Gamma(1-i\xi)}( 1 - \frac{i\xi^2}{z} +
\frac{\xi^2(1 + i \xi)^2}{2 z^2}) - \frac{i z^{i\xi}e^{iz}}{z
\Gamma(i\xi)}];
$$

$$
F_1 = F ( i\xi +1, 2, i z) \simeq  \frac{ i e^{-\pi\xi/2}}{z}[
\frac{z^{-i\xi}}{\Gamma(1-i\xi)}( 1 - \frac{\xi(1 + i \xi)}{z}) -
\frac{ z^{i\xi}e^{iz}}{ \Gamma(1 + i\xi)}( 1 - \frac{\xi(1 - i
\xi)}{z})];
$$

In the result the leading term in the left side of Eq. (\ref{36})
is the following

\begin{eqnarray}
\label{39} J_1 \simeq  e^{-\pi\xi} \frac{\sinh \pi \xi}{\pi \xi}\{2k r - \nonumber\\
\frac{1}{ k r} [ \xi^2 +  \Re \{\frac{\Gamma(1+i\xi)e^{-2i k r - 2
i \xi \ln 2 k r}}{\Gamma(-i\xi)}\}]\} +  O [\frac{1}{(k r)^2}].
\end{eqnarray}

This value defines  variation of the flux directed along the
incident wave vector and it grows linearly together with the
distance from the scattering center analogously to the total cross
section. As it was mentioned  above (Eq. (\ref{34})), this growth
is not connected with increase of the particle flux but describes
distorted part of the wave front which is extended together with
$r$ because of long-range character of the potential.

Calculation of the integral
$$
J_2 =\frac{1}{ k r} \int_{0}^{2 k r}|F ( i\xi, 1, i z)|^2 z d z
$$

by means of the analogous technique leads to the following result

\begin{eqnarray}
\label{40} J_2 \simeq  e^{-\pi\xi} \frac{\sinh \pi \xi}{\pi
\xi}\{2k r + \frac{2}{ k r} [ \xi^2(- \frac{1}{2} -
\Re \psi(1 + i\xi) + \ln 2 k r) - \nonumber\\
\Re \{\frac{\Gamma(1+i\xi)e^{-2i k r - 2 i \xi \ln 2 k
r}}{\Gamma(-i\xi)}\}]\} + O [\frac{1}{(k r)^2}],
\end{eqnarray}

\noindent where  $\psi(x)$ is the logarithmic derivative of
$\Gamma$ - function \cite{mors}.

The last integral in Eq. (\ref{36})

\begin{eqnarray}
\label{41} J_3 = \frac{\xi}{k r}\int_{0}^{2 k r} z \Im [ F(-i
\xi, 1, - i z) F(i \xi + 1, 2, i z) ] d z =\nonumber\\
-\frac{\xi}{k r}\int_{0}^{2 k r}  \Re [ F(-i \xi, 1, - i z) F(i
\xi + 1, 1, i z) ] d z,
\end{eqnarray}

transforms as follows

\begin{eqnarray}
\label{42} J_3 \simeq  - \frac{2}{ k r} e^{-\pi\xi} \frac{\sinh
\pi \xi}{\pi \xi}\{ \xi^2( - \Re
\psi(1 + i\xi) + \ln 2 k r) - \nonumber\\
\frac{1}{2}\Re [\frac{\Gamma(1+i\xi)e^{-2i k r - 2 i \xi \ln 2 k
r}}{\Gamma(-i\xi)}]\} + O [\frac{1}{(k r)^2}].
\end{eqnarray}

Substitution of Eqs. (\ref{39}) - (\ref{42}) to Eq. (\ref{36})
shows that it is satisfied with the considered accuracy. Besides,
one can see that the left side of Eq. (\ref{36}) coincides with
the total cross section (\ref{23}) in the limit $k r \gg 1$, and
the right side of Eq. (\ref{36}) transforms to the imaginary part
of the scattering operator (\ref{36}) with $\vec n = \vec{n}'$.
So, we can consider this calculation as the proof of the "optical"
theorem for the Coulomb scattering problem.

\section{Movement of the wave packet in the Coulomb field}

As it follows from the results of the preceding sections,
regularization of the Rutherford cross section is defined by the
characteristic angle

\begin{eqnarray}
\label{43} \theta_0 = \sqrt{\frac{2}{kr}},
\end{eqnarray}

\noindent which corresponds to the boundary of "near" zone and is
considered as the kinematic parameter (KP) of the system. However,
in real scattering experiments the incident particle is actually
represented by the localized wave packet  \cite{goldberger}.
Besides, the Coulomb potential is screened at some distance $R_s$,
depending on the properties of the medium where the collision is
happened. Therefore in general case the problem is characterized
by some additional parameters that can be considered as the
external parameters (EP). So, it is essential to estimate the
conditions when the KP is more important for the cross section
regularization that the EP. We will take into account two the most
essential EP: the screening angle $\theta_s$ the incident angle
parameter  $\theta_{int}$, depending on the wave packet
transversal width a and defining the zone of interference between
the incident and scattered waves (see also Sec.2).  The simple
estimation of these parameters leads to

\begin{eqnarray}
\label{44} \theta_s = \frac{1}{k R_s}, \qquad \theta_{int} =
\frac{a}{r}.
\end{eqnarray}

Evidently, the kinematic regularization is the most essential if
the angle width of the near zone is larger in comparison with the
characteristic angle intervals connected with  EP, that is the
following conditions are fulfilled

\begin{eqnarray}
\label{45}\theta_0 > \theta_s,\quad \frac{k R^2_s}{r} > 1; \qquad
\theta_0 > \theta_{int}, \quad \frac{k a^2}{r} < 1.
\end{eqnarray}

The first inequality depends on the mechanism of screening and
should be analyzed for every concrete system as it will be
considered below (Sec.6) for the scattering by impurities in
semiconductors. In order to take into account the finite size of
the wave packet in the second inequality in (\ref{45}) one should
use the time-dependent theory of collisions  \cite{goldberger},
\cite{1971}, that we will consider in this section.

Let us suppose that the initial state of the particle in the moment $t = 0$ is defined by the wave packet in the
following form

\begin{eqnarray}
\label{46}\Psi_{\vec k} (\vec r, 0) = \int d \vec q\Phi(\vec q -
\vec k)e^{i \vec q (\vec
r - \vec r_0)} \equiv  e^{i \vec k (\vec r - \vec r_0)} G (|\vec r - \vec r_0|),  \nonumber\\
G (\rho) = \int d \vec p\Phi(\vec p)e^{i \vec p \vec{\rho}},
\end{eqnarray}

\noindent where  $\vec r_0$ is the coordinate corresponding to the
initial position of the wave packet ; $\Phi(\vec p)$ are the
amplitudes of probabilities of the wave vector distribution near
the center $\vec k$ in the initial state; $G (\rho)$ is the
function which describes the form of the localized wave packet in
the coordinate space \cite{goldberger}.

In order to describe evolution of the wave packet (\ref{46}) it
should be expanded in the solutions of the stationary
Schr\"odinger equation with the Coulomb potential  \cite{1971}
(let us consider the attractive potential for the definiteness )

$$
\psi_{\vec{k}}(\vec{r}) =  N  e^{i \vec k \vec r} F [ i\xi_k, 1, i
(k r - \vec{k}\vec{r})].
$$

In the standard experimental setting  (Fig.5) the initial position
of the wave packet corresponds to the condition $z_0 \rightarrow -
\infty$. In this case the stationary wave function
$\psi_{\vec{k}}(\vec{r})$ coincides with the  plane wave
\cite{goldberger} and the  expansion of  $\Psi_{\vec k} (\vec r,
0)$ in the functions $\psi_{\vec{k}}(\vec{r})$ includes the same
coefficients as in the formula (\ref{46}) with an accuracy to the
terms of the order  $|z_0|^{-1}$ conditioned by the logarithmic
distortion of the wave front in the Coulomb field \cite{landau1}.
In the result the wave function describing the wave packet state
in an arbitrary moment of time has the following form

\begin{eqnarray}
\label{47}\Psi_{\vec k} (\vec r, t) = \int d \vec q\Phi(\vec q -
\vec k)e^{i \vec q \vec r}e^{-i \vec q \vec r_0} e^{ \pi\xi_q/2}
\Gamma (1 - i\xi_q) F [ i\xi_q, 1, i (q r -
\vec{q}\vec{r})]e^{-i\hbar q^2 t/2m}.
\end{eqnarray}

As it was investigated in detail in the monography
\cite{goldberger} the wave packet spread (diffraction) can be
neglected during time of the  interaction in real scattering
experiments. This corresponds to the following approximations in
the integrand expression in the formula (\ref{47})

\begin{eqnarray}
\label{48} \vec q - \vec k = \vec p;\quad p \ll k;\quad
\frac{\hbar q^2}{2 m}\simeq \frac{\hbar k^2}{2 m} + (\vec p\vec
v);\quad \xi_q \simeq \xi = \frac{\alpha }{\hbar v};\quad q \simeq
k + \frac{(\vec p\vec k)}{k},
\end{eqnarray}
\noindent where $\vec v = \hbar \vec k/m$ is the group velocity of
the center of the wave packet coinciding with the velocity of
classical particles.

Let us remind briefly results of the time-dependent collision
theory in the case of the short-range potential when the
asymptotic form  (\ref{5}) of the stationary wave function can be
used for analysis of the wave packet evolution  \cite{goldberger}

\begin{eqnarray}
\label{49}\Psi_{\vec k} (\vec r, t) = \int d \vec q\Phi(\vec q -
\vec k)e^{-i \vec q \vec{r}_0} [ e^{i \vec{q}\vec{r}} +
f(\theta_q) \frac{e^{iqr}}{r}]e^{-i\hbar q^2 t/2m},
\end{eqnarray}
\noindent where  $\theta_q$ is the angle between the vectors $\vec
q$ and $q\vec r/r$.

Now one can use the expansions  (\ref{48}) and to find the
following result for the function $\Psi_{\vec k} (\vec r, t)$

\begin{eqnarray}
\label{50}\Psi_{\vec k} (\vec r, t) \simeq  \{e^{i \vec k \vec r }
G (|\vec r - \vec r_0 - \vec v t|) + f(\theta_k)\frac{e^{ikr}}{r}G
(|r \frac{\vec k}{k} - \vec r_0 - \vec v t|)\}e^{-i \vec k \vec
r_0 }e^{-i\hbar k^2 t/2m}.
\end{eqnarray}

Fig.5 shows the sketch of distribution of the probability density
corresponding to the wave packet (\ref{50}) in some moment t. It
demonstrates two essential results which represents actually the
basis for use the quantum mechanical stationary scattering theory
for  description of the collisions between real particles
\cite{goldberger}. Firstly, the overlapping of the fluxes
corresponding to the incident ( first term in the formula
(\ref{50}))  and scattering particles is essential only in the
above-mentioned interference zone with the angular width
$\theta_{int} = a/r$ and they can be considered separately out of
this domain. Besides, the scattering flux is localized in the
spherical layer with the average radius  $r \simeq |\vec r_0 +
\vec v t|$ and width  $\sim a$. The angular distribution of the
scattering particle in the limits of this layer is completely
defined by the scattering amplitude  $f(\theta_k)$ calculated on
the basis of the stationary theory.

\begin{figure} [h]
\includegraphics[scale=0.8]{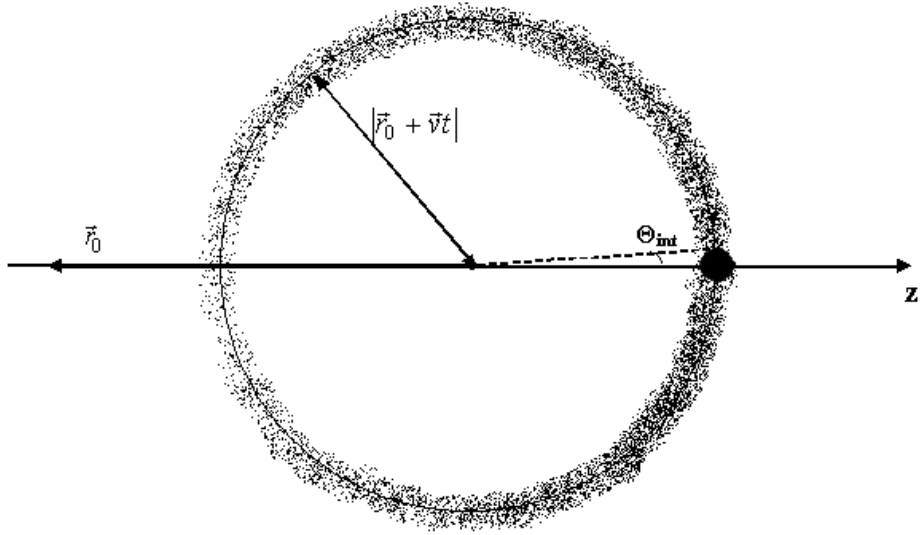}%width = 14 cm,height =9 cm
 \caption{ Sketch of distribution of the probability density
corresponding to the wave packet (\ref{50}) in some moment t}
\end{figure}

The expansions (\ref{48}) can be used in the integral  (\ref{49})
in the case of the integrand without singularities in the range of
the variable variation . This condition doesn't satisfied for the
asymptotic form  (\ref{5}) in the case of the Coulomb field
because the Rutherford amplitude includes unintegrable
singularity. Let us show, however, that the representation of the
wave packet analogous to the formula (\ref{50}) is justified also
for the Coulomb problem if the expansion  (\ref{50}) is built on
the basis nonasymptotic representation  (\ref{11a}) for the
confluent hypergeometric function:

\begin{eqnarray}
\label{51}\Psi_{\vec k} (\vec r, t) = \int  d \vec q \Phi (\vec q
- \vec k)e^{-i \vec q \vec{r}_0} [(z_q)^{i \xi_q}\frac{e^{i q
r}}{\Gamma (i \xi_q)} G_1(i\xi_q, i z_q) + \nonumber\\ (z_q)^{- i
\xi_q}\frac{e^{i \vec q \vec{r}}}{ \Gamma (i \xi_q)} G_2( i\xi_q,
i z_q)]e^{-i\hbar q^2 t/2m};
\nonumber\\
z_q = q r - \vec{q}\vec{r}.
\end{eqnarray}

The functions $G_{1,2}$ are rather smooth and integrable. One can
use the expansion (\ref{48}) for their arguments if the following
condition is satisfied in the region $z_q \leq 1$ of the most
essential variation of these functions

\begin{eqnarray}
\label{52} z_k \geq \vec p ( r \frac{\vec k}{k} - \vec r )\simeq p
r \theta_{p} .
\end{eqnarray}

If the spread of the wave packet is neglected, the value   $|\vec
p|$ can be estimated as $|\vec p|\simeq k \theta_{p} \simeq k a
/r$ (à is the characteristic linear size of the wave packet
localization in space) and the condition (\ref{52}) leads to the
inequality

\begin{eqnarray}
\label{53} \theta^2 \simeq \frac{1}{k r}  \geq (\frac{a}{r})^{2},
\quad \frac{k a^2}{r}\leq 1.
\end{eqnarray}

\noindent It coincides with the above mentioned estimation
(\ref{45}) considered on the basis of the qualitative analysis.

In the result the functions  $G_{1,2}$ in the formula (\ref{51})
can be removed out of the integral with the arguments
corresponding to the center of the wave packet and it leads to the
expression

\begin{eqnarray}
\label{54}\Psi_{\vec k} (\vec r, t) = [(z_k)^{i \xi_k}\frac{e^{i k
r}}{\Gamma (i \xi_k)} G_1(i\xi_k, i z_k)G (|r \frac{\vec k}{k} -
\vec r_0 - \vec v t|) + \nonumber\\ (z_k)^{- i \xi_k}\frac{e^{i
\vec k \vec{r}}}{ \Gamma (i \xi_k)} G_2( i\xi_k, i z_k)G (|\vec r
- \vec r_0 - \vec v t|)]e^{-i \vec k \vec{r}_0}e^{-i\hbar k^2
t/2m};
\nonumber\\
z_k = k r - \vec{k}\vec{r}.
\end{eqnarray}

It means that the scattering process in the Coulomb field can be considered on the basis of the stationary
theory as it takes place in the case of the short-range potential. Besides, the  incident and scattered wave
packets are extending in the space separately excluding unessential domain of their overlapping.

\section{Calculation of the charge carrier mobility in the extrinsic semiconductors }

It is important to consider the concrete physical system where the
described peculiarities of the scattering process in the Coulomb
field can be appeared for some observed characteristics.
Accordingly to the estimation  (\ref{45}), it is possible if the
following inequality is fulfilled

\begin{eqnarray}
\label{55} \frac{k R^2_s}{r} > 1.
\end{eqnarray}

\noindent Here $R_s$ is the screening radius of the Coulomb
potential in a medium and it depends on the screening mechanism in
the system. The value r is defined by the distance between the
scattering center and detector or by the average distance between
two subsequent collisions if the scattering operator (\ref{26}) is
used for the description of kinetic processes in the system.

In the present paper the nonasymptotic scattering theory will be
used for analysis of the charge carrier mobility in the extrinsic
semiconductors for low temperature. In this case  concentration of
the impurity centers defines both the type of the carriers and
their concentration and also the main contribution to the
resistance of the semiconductor \cite{ziman}. The problem was
recently analyzed in detail in the paper \cite{poklonskii} and
results of the various phenomenological models for regularization
of the Rutherford cross-section were compared with the
experimental data \cite{poklonskii}. It was shown that the wide
used models of Brooks-Herring \cite{herring}, and
Conwell-Weisskopf \cite{conwell} don't describe completely the
experimental dependence of the mobility on the temperature and
impurity concentration. The authors of the paper \cite{poklonskii}
fitted the experimental data essentially better by means of an
additional phenomenological parameter with the physical meaning of
the characteristic time of the collision. It seems to us that such
parameter takes into account partly the influence of the "near"
zone (see Sec.2) on the formation of the scattered flux. So, the
regularization of the scattering problem in the Coulomb field is
of interest not only as the methodical problem but also as the
applied one.

Let us consider the extrinsic semiconductor with the concentrations of the donors  $n_1$ and acceptors $n_2$ in
the charge states $Z_1e $ and  $ Z_2e$ correspondingly (in the most of real structures the impurities with the
charge $|Z_{1,2}| = 1$ are mainly important ), e is the absolute value of the electron charge.

In general case the value $n_e$ is defined by both the thermally
excited carriers and the carriers conditioned by the impurities.
The semiconductors with the wide forbidden zone were analyzed in
the paper \cite{poklonskii} and the value $n_e$ can be estimated
as

$$
n_e \simeq Z_{1}n_{1}- Z_{2}n_{2}= n,
$$

\noindent for the considered low temperature.

Let us introduce also another parameter which is more spread in
the semiconductor physics:   K is the compensation and is usually
a quite small value  $K\simeq 0.1$ \cite{poklonskii}

\begin{eqnarray}
\label{56}  n_1 = \frac{n}{Z_{1} - K Z_{2}}, \quad n_2 = \frac{n
K}{Z_{1} - K Z_{2}}, \quad K = \frac{n_2}{n_1}.
\end{eqnarray}

It is well known \cite{ziman} that the Coulomb potential screening
in semiconductors is defined by several factors. From one side,
there is the static dielectric constant $\epsilon$ conditioned by
the electrons from the valency zone which doesn't change the
long-range character of the potential. From the other side, the
Debye screening of the potential by free electrons (or holes)
leads to its cut off on the distance \cite{ziman}

\begin{eqnarray}
\label{57} R_s \simeq \sqrt{\frac{\epsilon k_B T}{4 \pi e^2 n_e}},
\end{eqnarray}

\noindent where $k_B$ is the Boltzmann constant; $T$ is the
crystal temperature; $n_e$ is the concentration of free charge
carriers (electrons in the conductivity zone for n-type
semiconductors or holes in the valency zone for p-type
semiconductors).

The average distance r between scattering centers and the
characteristic wave vector for the carriers in the formula
(\ref{55}) can be estimated as
$$
r \simeq n^{-1/3}; \qquad k = \frac{\sqrt{2 m^* E}}{\hbar} \simeq
\frac{\sqrt{3 m^* k_B T}}{\hbar},
$$
\noindent with $m^*$ as the carrier effective mass.

In the result the condition (\ref{55}) leads to the following
inequality
\begin{eqnarray}
\label{58}  \frac{(3\epsilon m^*)^{1/2} (k_B T)^{3/2}}{4 \pi e^2
\hbar n^{2/3}} > 1,
\end{eqnarray}

\noindent which is fulfilled in the entire range of the density
and temperature considered in \cite{poklonskii}.

In the most applications the theoretical estimation of the carrier
mobility is based on the approximation of relaxation time  $\tau$
and the Maxwell velocity distribution. It leads to the following
formula (n-type semiconductors are considered for the
definiteness) \cite{ziman}

 \begin{eqnarray}
 \label{59}
\mu = \frac{e}{m^*} <\tau>
\end{eqnarray}
\begin{eqnarray}
\label{60}<\tau> = [\int_0^{\infty} E^{3/2} e^{- E/k_B T}]^{-1}
\int_0^{\infty}  \tau(E) E^{3/2} e^{- E/k_BT}.
\end{eqnarray}

Here the relaxation time is supposed to be averaged on the energy of carriers with the Maxwell distribution.

It is known \cite{blekmor} that if the several mechanisms of
scattering take place (in our case there are scattering by donors
and acceptors), the more accurate result the additional averaging
on the types of scattering centers should be fulfilled:

\begin{eqnarray}
\label{61}\ \tau(E) = \frac{\tau_1(E) \tau_2(E)}{\tau_1(E) +
\tau_2(E)},  \nonumber\\
\tau_{1,2}(E) = \frac{1}{n_{1,2} v \sigma_{tr1,2}},
\end{eqnarray}

\noindent where the indexes 1,2 correspond to the scattering by donors and acceptors;  $\sigma_{tr1,2}$ is the
transport cross-section for the cases of the attraction and repulsion. In accordance with Sec.3 these values are
defined by the formulas :

\begin{eqnarray}
\label{62} \sigma_{tr1,2} = \frac{2\pi \xi_{1,2}^2 e^{\mp
\pi\xi_{1,2}}}{k^2} \int_{0}^{2 k r_{1,2}}|U ( 1\pm i \xi_{1,2},1,
i z)|^2 z d z .
\end{eqnarray}

We use here the more accurate formula than Eq.(\ref{27}) because
in this case the condition  $kr\gg 1$ can not be fulfilled.

The parameters of interaction between carriers and scattering centers in the considered cases are the following

$$
\xi_{1,2}(E) = \frac{Z_{1,2} e^2}{\epsilon \hbar v}
$$
\noindent and  the static dielectric constant of the crystal is taken into account.

Accordingly to the formulae (\ref{62}) the transport cross section
depends on the potential charge as distinct of its calculation
with the Rutherford cross section. The similar effect ("phase
shift") is well known for extrinsic semiconductors and considers
usually by means of the Fridel sum rule \cite{fridel}. Indefinite
parameter $r$ is included in Eq.(\ref{62}). If the value $\mu$ is
calculated by the totally microscopic way it should be averaged on
the space distribution of the impurities in the sample. It is
equivalent to the integration of the expression (\ref{59}) by $r$
taking into account Eq.(\ref{62}). However, the transport
cross-section has the smooth logarithmic behavior on $r$ which can
substituted in Eq.(\ref{62}) as the average distance between the
impurities with the considered accuracy. Then the value $r=0.5
n^{-1/3}_i$ can be used in Eq.(\ref{62}) analogously to the both
models  \cite{herring}, and   \cite{conwell}.

It is convenient to define the auxiliary value  $\sigma_{tr}'$ so, that
$$\sigma_{tr}= 2 \pi\frac{\xi^{2}}{k^{2}} \sigma_{tr}'.$$  Then nonasymptotic calculation leads to

\begin{eqnarray}
\label{63} \sigma_{tr1,2}' =  e^{\mp \pi\xi_{1,2}} \int_{0}^{2 k
r_{1,2}}|U( 1\pm i \xi_{1,2},1, i z)|^2 z d z ,
\end{eqnarray}

with the values

$$
r_{1}=\frac{(Z_{1}-K Z_{2})^{1/3}}{2 n^{1/3}} , \quad r_{2}=\frac{(Z_{1}-K Z_{2})^{1/3}}{2(n K)^{1/3}},
$$

which are defined by the half of the average distance between the donors and acceptors correspondingly.

In the result the following expression for the carrier mobility can be obtained:

\begin{eqnarray}
\label{64} \mu = \frac{2^{5/2}\epsilon^2 (k_B T)^{3/2} }{3
\pi^{3/2}  e^3 m^{\ast1/2} n} \int_0^{\infty}\frac{x^{3} e^{-
x}}{Z_{1}^{2}\sigma_{tr1}'(x)+ K Z_{2}^{2} \sigma_{tr2}'(x)}d x.
\end{eqnarray}

The integrals over energies can be estimated by the standard way
\cite{conwell}: the smoothly changing functions can be taken out
of the integrals with argument $x=3$ when the energy distribution
function has the maximum value. It leads to the following
analytical expression for the mobility:

\begin{eqnarray}
\label{65} \mu = \frac{2^{7/2}\epsilon^2(Z_{1}-K Z_{2}) (k_B
T)^{3/2} }{ \pi^{3/2}   e^3 m^{\ast1/2} n
(Z_{1}^{2}\sigma_{tr1}'(3k_B T)+ K Z_{2}^{2} \sigma_{tr2}'(3k_B
T))}.
\end{eqnarray}

In the case $Z_{1}=Z_{2}=1$ it transforms as follows

\begin{eqnarray}
\label{66} \mu = \frac{2^{7/2}\epsilon^2(1-K) (k_B T)^{3/2} }{
\pi^{3/2}   e^3 m^{\ast1/2} n (\sigma_{tr1}'(3k_B T)+ K
\sigma_{tr2}'(3k_B T))}.
\end{eqnarray}

We can compare it with the analogous formula in the framework of
the Conwell-Weisskopf model \cite{conwell}

\begin{eqnarray}
\label{20b}\mu_{CW} = \frac{2^{7/2}\epsilon^2 (1-K) (k_B T)^{3/2} }{ \pi^{3/2}   e^3 m^{\ast1/2}n (1+K)
\ln(1+(\frac{3\epsilon k_B T (1-K)^{1/3}}{Z e^{2}(n (1+K))^{1/3}})^{2})}.
\end{eqnarray}

The results of calculation by means of Eqs.(\ref{66}) and their
comparison with  the Conwell-Weisskopf model results are shown in
Fig.6. The same figure shows that the dependence of the mobility
on the temperature and compensation K in our consideration differ
essentially on the results of the Conwell-Weisskopf model
\cite{conwell} based on the Rutherford cross section with the
phenomenological regularization. In principle, such distinction
can be discovered in some experiments.

%\newpage
\begin{figure} [h]
%\begin{minipage}{0.2\texwidth}\centering
\includegraphics[width = 5cm,height =5cm]{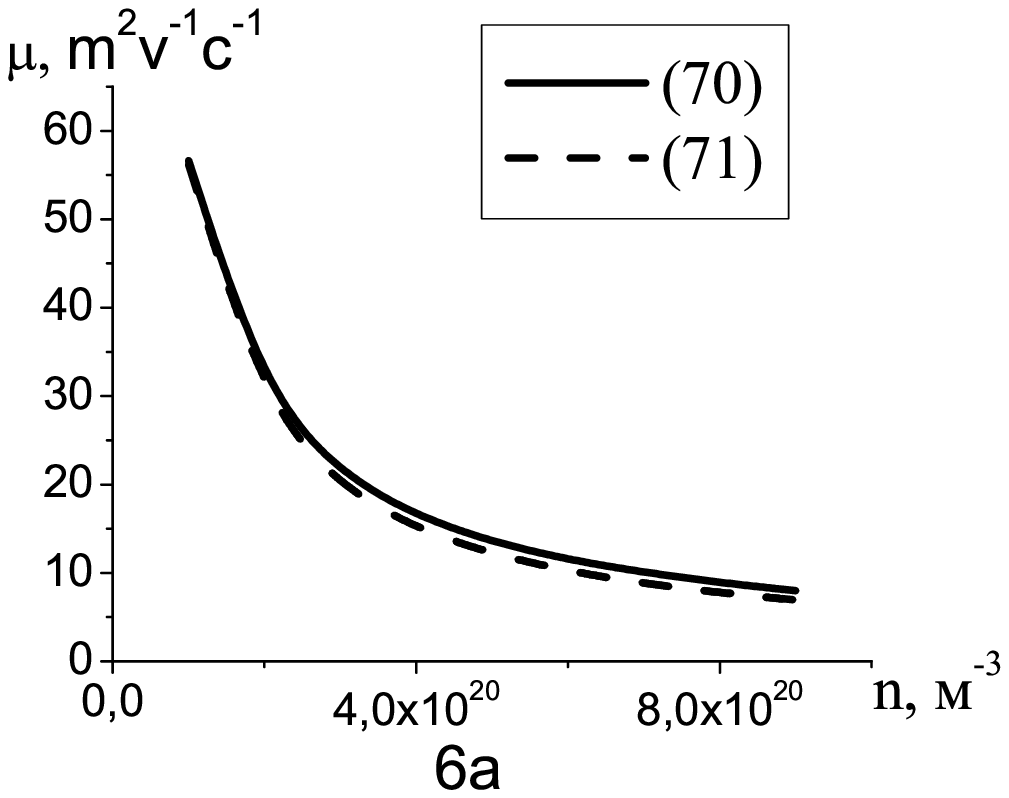}
%\end{minipage}\hfill\
%\begin{minipage}{0.2\texwidth}\centering
\includegraphics[width = 5cm,height =5cm]{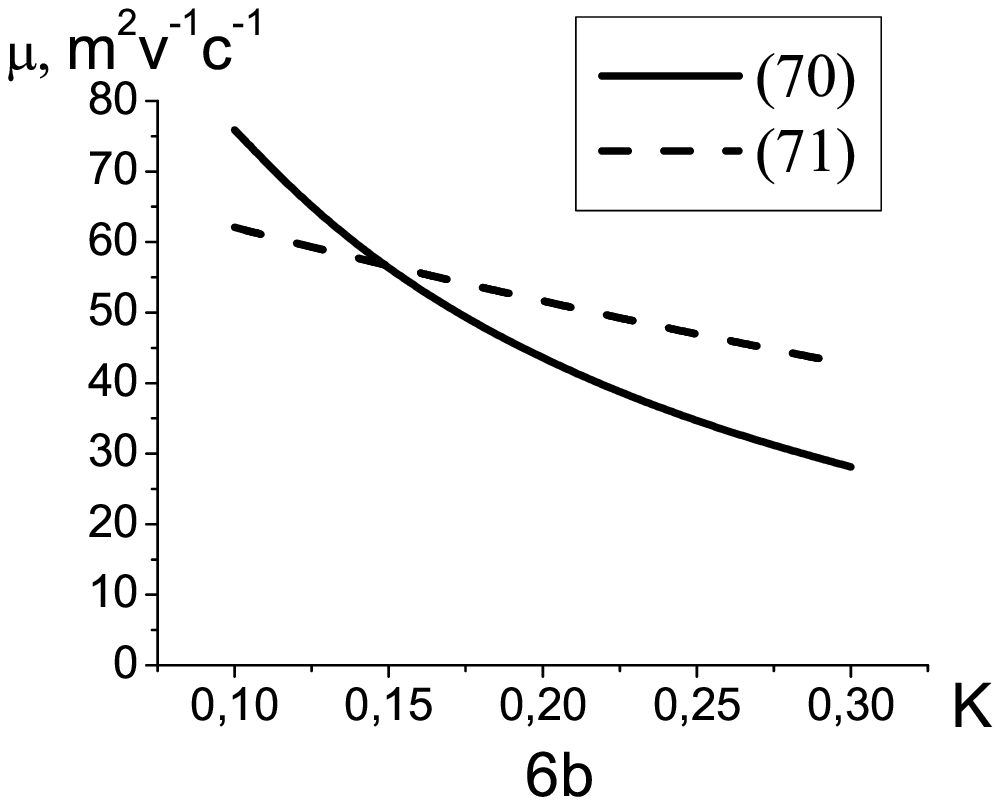}
%\end{minipage}\hfill\
%\begin{minipage}{0.2\texwidth}\centering
\includegraphics[width = 5cm,height =5cm]{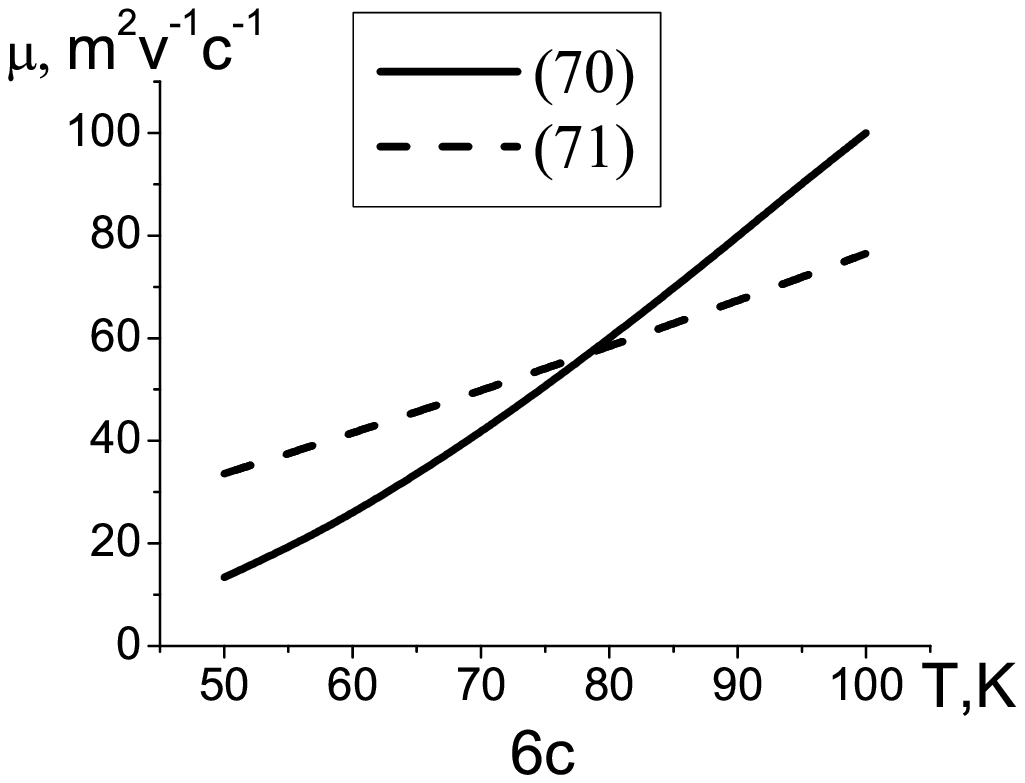}
%\end{minipage}\hfill\
\caption{ Comparison of the mobilities calculated with the
nonasymptotic transport cross-section (solid line) and in the
framework of the Conwell-Weisskopf model (dashed line)(6a -
dependence on the impurity concentration; 6b - on the
compensation; 6c - on the temperature). The following parameters
were used $ T = 78Ê, \epsilon = 10, m = 0.2 m_0 , K = 0.15$.  }
\end{figure}

\section{Acknowledgments}

Authors  are grateful to Prof. N.A.Poklonskii for useful
discussions and International Scientific Technical Center (Grant
B-626) for the support of this work .

%\listoffigures


\begin{thebibliography}{}

\bibitem{landau1} L.D.Landau and E.M.Lifshitz, {\it Quantum Mechanics:
Non-Relativistic Theory}, 3rd edition Vol 3, (Pergamon Press,
London, 1997).

\bibitem{newton} R.G.Newton, {\it Theory of Waves and Particles}, 2nd edition , (McGraw Hill, New York, 1982).

\bibitem{landau2} L.D.Landau and E.M.Lifshitz, {\it Mechanics}, (Nauka, Moscow, 1965)

\bibitem{sing} D.Sing, {\it Classical Dynamics},( Fizmatgiz, Moscow, 1963).


\bibitem{herring} H.Brooks,  {\it Phys. Rev.},
{\bf 83},(1951), 879.

\bibitem{conwell} E.Conwell and V.F.Wesskopf,  {\it Phys. Rev.},
{\bf 77},(1950), 388.

\bibitem{poklonskii} N.A.Poklonskii, S.A.Vyrko et al.,{\it Applied Phys.},
{\bf 93},(2003),  9749.

\bibitem{semiconducter} B.K.Ridley, {\it Quantum Processes in Semicoducters}, (Clarendon Press, Oxford, 1999);
K.Seeger, {\it Semiconductor Physics}, (Springer-Verlag, Berlin,
1999).

\bibitem{wire} S.W.Kim, H-K.Park, H-S. Sim and H.Shomerus,{\it J.Phys.A: Math. and Gen.},
{\bf 36},(2003), 1299.

\bibitem{film} K.Elmer,{\it J. Phys.D: Applied Phys.}, {\bf
34},(2001), 3097.

\bibitem{tube} K. Harigawa,{\it J.Phys: Condensed Matter}, {\bf 12},(2000), 7069.
388.

\bibitem{1971} V.G.Baryshevskii, L.N.Korennaya and I.D.Feranchuk,{\it Soviet Physics JETP}, {\bf 34},(1972), 249.

\bibitem{goldberger} M.Goldberger and K.Watson, {\it Collision Theory},(Wiley, New York, 1964).


\bibitem{zack} W.Zackowicz,{\it J.Phys.A: Math. Gen.},
{\bf 36},(2003), 4445.

\bibitem{born} J.Jackson , {\it Classical electrodynamics}, (John Willey and sons, New-York - London, 1962).

\bibitem{mors} Ph.Mors and H.Feshbach {\it Methods of Theoretical Physics},
(Mc-Graw-Hill Book Co., New York, 1953).

\bibitem{ziman} J.M.Ziman {\it Principles of The Theory of Solids},
(At The University Press, Cambridge, 1972).

\bibitem{fridel} A.D.Boardman, D.W.Henry  {\it Phys. stat., sol. (b)},{\bf 60},(1973), 633.

\bibitem{blekmor} J. Blekmor {\it  Solid state physics}, (Nauka, Moscow,
1988).

\bibitem{kinetic} E.M.Lifshitz, L.P. Pitaevski {\it Physical kinetics }, Moscow,1979.

\bibitem{nikiforov}A.F. Nikiforov, V.B.Uvarov {\it  Special functions of mathematical physics}, (Nauka, Moscow,
1984).

\end{thebibliography}
\end{document}